\documentclass[conference]{IEEEtran}
\usepackage{amsmath, amssymb, amsfonts}
\usepackage{graphicx}
\usepackage{cite} 
\usepackage{textcomp}
\usepackage{xcolor}
\usepackage[utf8]{inputenc}
\usepackage{algorithm}
\usepackage{algorithmic}
\usepackage[hidelinks,bookmarks=false]{hyperref}
\usepackage[T1]{fontenc}
\begin{document}


\title{Uncertainty-Aware Haptic Signal Estimation for Reliable and Resource Efficient Tactile Internet}

\author{\IEEEauthorblockN{Georgios Kokkinis\textsuperscript{1},
Alexandros~Iosifidis\textsuperscript{2}, and Qi Zhang\textsuperscript{1}}
\IEEEauthorblockA{\textsuperscript{1}\textit{DIGIT and Dept. of Electrical and Computer Engineering,}
\textit{Aarhus University, Denmark}\\
\IEEEauthorblockA{\textsuperscript{2}\textit{Data Science Research Centre, Tampere University, Tampere, Finland}} 
Email: \{gkokkinis, qz\}@ece.au.dk; 
alexandros.iosifidis@tuni.fi}
}

\maketitle

\begin{abstract}
The Tactile Internet aims to enable real-time remote haptic interaction; however, the high sampling rates required for transparency in haptic control often lead to severe congestion in multi-user wireless environments. This paper proposes the Agile AI-empowered Haptic (A$^2$HAP) framework, which integrates VarxHAP, a novel probabilistic neural network for joint force and uncertainty estimation, with an error-resilient controller. By employing a hierarchical gating architecture, the system dynamically adapts transmission thresholds to balance model confidence against reliability targets. Simulation results demonstrate that A$^2$HAP suppresses packet rates by up to 45\% during peak traffic and reduces resource block consumption by 25\% on average. Consequently, the framework supports a 20\% increase in user capacity compared to state-of-the-art methods while maintaining the ultra-reliability required for stable teleoperation.
\end{abstract}


\section{Introduction}
\label{sec:Intro}
The Tactile Internet (TI) introduces a paradigm shift in telecommunications, aiming to enable the remote delivery of physical touch and real-time haptic interaction. Unlike traditional communication, haptic teleoperation requires the transmission of kinesthetic data within a closed-loop control system. To maintain ``perceptual transparency" and control stability between a human operator and a remote robot, the network must satisfy stringent requirements, such as sub-millisecond latency and extreme reliability, often exceeding $99.999\%$~\cite{antonakoglou2018}.

The primary bottleneck in realizing large-scale haptic systems is the high sampling rate required (i.e., typically 1 kHz) to provide smooth force feedback. In multi-user wireless environments, transmitting at such high packet rates leads to severe network congestion, increased interference, and frequent deadline violations. Traditional traffic reduction techniques, such as Perceptual Deadband (PD) filtering successfully suppress packets that are not perceptible by users~\cite{Hinter2008}. However, PD-based traffic reduction is high when human motion is low, but severe traffic spikes occur during high-motion activity.

Addressing burstiness in traffic remains a primary challenge in haptic communication. Research on this field includes adaptive velocity deadband coding~\cite{Kammer2010}, yet perceptual coding still struggles during high activity. Adaptive rate control schemes attempt to mitigate this by scaling thresholds based on network intensity~\cite{Gui2020}, often utilizing passivity-based control to maintain stability~\cite{Xu2019}. Additionally, edge-intelligent frameworks use reinforcement learning to optimize resource allocation under varying conditions~\cite{kokkinisDRL}. 

Beyond rate adaptation, recent literature explores predictive modeling to handle packet loss and complex dynamics. LSTM networks and model-mediated schemes manage non-linearities~\cite{MadsOctree}, while combining Shapley values with Gaussian processes and discrete mode decomposition increases robustness~\cite{Ali1, Ali2}. Model-mediated schemes also reduce traffic by transmitting only essential parameter updates during manipulations~\cite{MadsDT}. However, these predictive models typically operate independently of the MAC-layer; they lack mechanisms to directly translate model confidence into real-time packet suppression decisions. Similarly, while NetLfD uses network-aware confidence weights to improve robot learning, it lacks mechanisms to reduce traffic or ensure real-time URLLC reliability~\cite{Basak2023}. 

To address the disconnect between predictive AI and communication network constraints, we propose the Agile AI-empowered Haptic Communication (A$^2$HAP) framework. A$^2$HAP enables AI-native suppression, avoiding transmissions when the receiver can accurately predict the haptic signal, while providing robust recovery to reconstruct missing values lost to channel errors or congestion. 

The framework relies on two core components. The first is VarxHAP~\cite{kokkinisxHAP}, a probabilistic neural network estimator that outputs an estimated force value alongside an uncertainty metric. The second is an error-resilient controller, which utilizes proportional integral closed-loop elements to dynamically switch between estimation and transmission based on real-time network congestion.

\begin{figure*}[!t]
\label{avg_mse}
 \centering
 \includegraphics[width=1.0\textwidth]{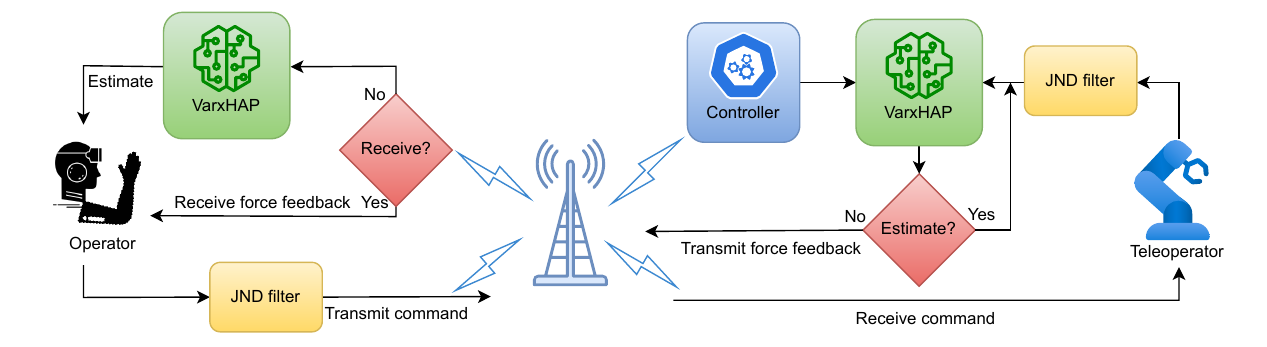}
 \caption{Overview of agile AI-empowered Haptic Communication (A$^2$HAP) framework.}
 \label{fig:A2HAP}
 \vspace{-1em}
\end{figure*}

The key contributions of this paper are summarized as follows:
\begin{itemize}
    \item \textbf{Uncertainty-Aware Haptic Estimator (VarxHAP):} A probabilistic model providing force estimates and calibrated uncertainty, enabling packet suppression and signal recovery.
    
    \item \textbf{Error-resilient Controller:} A closed-loop mechanism that dynamically scales a transmission threshold based on model uncertainty. It ensures the strict 99.999\% reliability target is maintained by determining precisely when to suppress physical packets in favor of local model inference.
    
    \item \textbf{AI-Empowered Framework (A$^2$HAP):} An agile framework for intelligent packet suppression and recovery. Instead of relying solely on reactive perceptual deadbands, it proactively replaces physical transmissions with confident model predictions. This approach accommodates 20\% more users than competing methods, achieving a transmission rate reduction ranging from 23\% in low traffic up to 45\% during peak traffic bursts.
\end{itemize}
For the remainder of the paper, the structure is as follows: Section~\ref{sec:Background} provides the theoretical foundation for human haptic perception, network performance constraints, and uncertainty-aware machine learning. Section~\ref{sec:A2HAP} describes the proposed framework, focusing on the logic of uncertainty-based transmission, uncertainty estimation, and the functions of the controller. Section~\ref{sec:varxhap} introduces the VarxHAP architecture, explaining how the model evaluates its own prediction confidence using a specialized training objective. Section~\ref{sec:results} presents simulation results to evaluate how the framework improves user capacity and resource efficiency compared to existing methods. Finally, Section~\ref{sec:conclusion} concludes the paper by summarizing key findings and discussing future directions for AI-driven haptic communications.
\section{Background}
\label{sec:Background}

This section provides the theoretical foundation for haptic perception and the constraints of Ultra-Reliable Low-Latency Communications (URLLC).
\subsection{Weber's Law and Haptic traffic}
\label{subsec:JND filter}
Human haptic perception is governed by the sensitivity of our biological receptors, as documented in~\cite{weber1996}. According to Weber’s Law, the JND is proportional to the intensity of the initial stimulus. In haptic teleoperation, this principle is utilized via Perceptual Deadband (PD) sampling. In PD coding, a transmission flag $f_{\text{JND}}$ is raised only if the deviation between the current sample $\mathbf{y}_{k,t}$ and the last successfully acknowledged reference $\mathbf{y}_{\text{ref}}$ exceeds a JND threshold $\delta = \kappa \cdot \|\mathbf{y}_{\text{ref}}\|$, where $\kappa$ denotes the Weber fraction. The flag is defined by:
\begin{equation}
    f_{\text{JND}} = \begin{cases} 
        1, & \text{if } \|\mathbf{y}_{k,t} - \mathbf{y}_{\text{ref}}\|_2 \ge \delta, \\ 
        0, & \text{otherwise}. 
    \end{cases}
\end{equation}
When $f_{\text{JND}} = 0$, the transmission is suppressed perceptually ($u_{k,t}=0$).
While PD effectively reduces average packet rates, rapid force transitions across multiple users can lead to synchronized traffic bursts, triggering network congestion.

\subsection{Network Requirements in URLLC}
Apart from congestion-induced errors, the primary challenges in realizing URLLC in mobile networks are the stringent reliability and latency targets. To ensure the stability of haptic teleoperation tasks, URLLC frameworks typically require a 1-millisecond end-to-end latency, though this may be relaxed to approximately 10 milliseconds depending on the specific control strategy and stability margins. Consequently, the window for packet retransmissions is extremely narrow. This restriction makes the ultra-reliability target of 99.999\% significantly more difficult to achieve, as the system has fewer opportunities to recover from channel-induced errors before the deadline of the haptic signal expires. 


\begin{table}[!t]
  \caption{Notation Summary}
  \label{tab:notation}
  \centering
  \renewcommand{\arraystretch}{1.15}
  \resizebox{\columnwidth}{!}{%
  \begin{tabular}{ll|ll}
    \hline
    \textbf{Symbol} & \textbf{Meaning} & \textbf{Symbol} & \textbf{Meaning} \\
    \hline
    $\mathcal{K}, K$ & User set and count & $\kappa$ & Weber fraction \\
    $\mathbf{y}_{k,t}, \hat{\mathbf{y}}_{k,t}$ & True and reconstructed force & $u_{k,t}$ & TX request indicator \\
    $\mathbf{H}_{k,t}$ & Received haptic history & $f_{\text{JND}}$ & Perceptual gate flag \\
    $\mathbf{O}_{k,t+m}$ & Updated operator intent & $f_{\text{AI}}$ & Model confidence flag \\
    $\hat{\sigma}_{k,t}$ & Mean estimation uncertainty & $\Psi_t$ & Control update direction \\
    $\mathbf{y}_{\text{ref}}$ & Last TX reference & $\delta$ & JND threshold \\
    $m$ & Consecutive estimation count & $R_t, R^*$ & Real and target reliability \\
    $\tau_k(m)$ & Adaptive uncertainty threshold & $e_t$ & Reliability error signal \\
    $\sigma_{b}$ & Base uncertainty limit & $E_{\text{est}}$ & Estimation-induced errors \\
    $\lambda$ & Threshold decay rate & $E_{\text{cong}}$ & Congestion-induced errors \\
    $K_p, K_i$ & PI control gains & $\Delta_t$ & Controller adjustment step \\
    $C$ & Total uplink capacity & $\eta$ & Base $\sigma$ scaling constant \\
    $P_{\text{bler}}$ & Wireless block error rate & $\mathcal{K}_{\text{req}}$ & Active request set \\
    $\delta_f$ & Noise floor threshold & $M_t$ & Masking indicator \\
    $\boldsymbol{\mu}_{k,t}$ & Point prediction (fallback) & $\mathbf{r}_t$ & Prediction residual vector \\
    $\mathcal{L}, \mathcal{L}_{(\cdot)}$ & VarxHAP total and component losses & $\omega_{\text{nll}}, \omega_{\text{cal}}$ & Loss weights \\
    $\mathbf{E}_m$ & Step-aware embedding & & \\
    \hline
  \end{tabular}%
  }
\end{table}

\section{Agile AI-empowered Haptic Communication (A$^2$HAP)}
\label{sec:A2HAP}

In this section, we propose our Agile AI-empowered Haptic Communication (A$^2$HAP) framework shown in Figure~\ref{fig:A2HAP}. We consider a haptic teleoperation scenario serving a set of users $\mathcal{K} = \{1, 2, \dots, K\}$ sharing a resource-constrained network. The system operates in discrete time steps $t$. Each application $k$ employs both a local and remote estimator that reduces bandwidth usage through uncertainty-based suppression. The A$^2$HAP framework integrates a perceptual JND filter described in Section~\ref{subsec:JND filter} with an estimator-controller loop described in Sections~\ref{subsec:Unc-Aware_Supress}--\ref{subsec:Ctrler} to minimize bandwidth usage, while ensuring targeted reliability.

\subsection{Uncertainty-Aware Suppression}
\label{subsec:Unc-Aware_Supress}
When $f_{\text{JND}} = 1$, i.e., the force change is perceptually noticeable, the system evaluates the uncertainty of the estimator. The adaptive uncertainty threshold $\tau_k(m)$ is defined as:
\begin{equation}
    \tau_k(m) = \sigma_{b} \cdot e^{-\lambda m},
\end{equation}
where $\sigma_b$ is the base uncertainty value and $m$ is the number of consecutive estimations. We exponentially decay the decision threshold as $m$ increases. This compensates for the compounding error in autoregressive estimation by forcing a ground-truth update before the accumulated drift becomes noticeable.
We define the model-based transmission flag $f_{\text{AI}}$ as:
\begin{equation}
    f_{\text{AI}} = \begin{cases} 
        1, & \text{if } \hat{\sigma}_{k,t} \ge \tau_k(m), \\ 
        0, & \text{otherwise}, 
    \end{cases}
\end{equation}
where $\hat{\sigma}_{k,t}$ is the uncertainty estimate.
The final transmission request $u_{k,t}$ is generated as the logical AND of the perceptual and model-based flags: $u_{k,t} = f_{\text{JND}} \land f_{\text{AI}}$.

\subsection{Probabilistic Haptic Estimation}
\label{subsec:hap_est}
VarxHAP operates on both sides of the network. Both the operator and the teleoperator run the estimator to continuously predict the current haptic force $\mathbf{y}_{k,t} \in \mathbb{R}^{3}$ based on the previously acknowledged history $\mathbf{H}_{k,t}$. When the estimation uncertainty is low, the sender safely suppresses physical transmissions, knowing the receiver's identical estimator will accurately reconstruct the force without disrupting communication.

The aggregated uncertainty estimate is calculated as the average standard deviation across all dimensions:
\begin{equation}
    \hat{\sigma}_{k,t} = \frac{1}{3} \sum_{d=1}^3 \exp\left(\frac{1}{2} \log v_{k,t,d}\right),
\end{equation}
where $\log v_{k,t,d}$ is the log-variance output of the estimator for dimension $d$ at time $t$. We give more details about the functions of the VarxHAP estimator in Section~\ref{sec:varxhap}, and provide the essential hyperparameter values in Table~\ref{tab:sim_params}.

\subsection{Error-Resilient Controller}
\label{subsec:Ctrler}
The uncertainty threshold $\tau_k(m)$ is adjusted by an adaptive controller. The controller parameters $\sigma_{b}$ and $\lambda$ are updated by a Proportional-Integral (PI) controller to meet a target reliability $R^*$. Let $R_t$ be the cumulative reliability observed at time $t$. The error signal is defined as $e_t = R^* - R_t$.
The controller distinguishes between errors due to estimation ($E_{\text{est}}$) and congestion ($E_{\text{cong}}$).

The direction of the update is determined by a signed term $\Psi_t$ that accounts for both the reliability state and the dominant error source:
\begin{equation}
    \Psi_t = \begin{cases} 
        -1, & \text{if } e_t < 0, \\ 
        \text{sgn}(E_{\text{est}} - E_{\text{cong}}), & \text{if } e_t \ge 0. 
    \end{cases}
\end{equation}

The magnitude of the update $\Delta_t$ is then computed as:

\begin{equation}
\label{eq:PI_ctrl}
    \Delta_t = \Biggl( K_p |e_t| + K_i \sum_{j=0}^t e_j \Biggr) \cdot \Psi_t.
\end{equation}
The second term on the right part of Eq.~\ref{eq:PI_ctrl} is the integral term and is used to eliminate steady-state errors, as the crucial point of the controller is to achieve the exact reliability target.
The parameters are updated according to:
\begin{equation}
    \lambda_{t+1} = \lambda_t + \Delta_t,
\end{equation}
\begin{equation}
    \sigma_{b, t+1} = \sigma_{b, t} - \eta \Delta_t,
\end{equation}
where $\eta$ is the adjusting weight to scale both controller parameters appropriately.

The control logic creates an asymmetric response based on the observed network state. When the system is under-utilized ($R_t > R^*$), the error $e_t$ is negative, causing the controller to decrease $\lambda$ and increase $\sigma_{b}$. This effectively raises the threshold $\tau_k(m)$, suppressing more packets to prioritize bandwidth efficiency. In cases of a reliability deficit ($R_t < R^*$), the error $e_t$ is positive, and the adjustment direction depends on the dominant error source. If the error is estimation-driven ($E_{\text{est}} > E_{\text{cong}}$), $\Delta_t > 0$, causing $\lambda$ to increase and $\sigma_b$ to decrease. This tightens the threshold and forces more frequent transmissions to refresh the local estimator. Conversely, if the error is congestion-driven ($E_{\text{cong}} > E_{\text{est}}$), $\Delta_t < 0$, causing the controller to relax thresholds. By reducing the packet injection rate, the system prevents a positive feedback loop where increased transmission requests would otherwise exacerbate network collapse.

\subsection{Channel Model and Reconstruction}
The network has a fixed uplink capacity $C$. If the number of requested users $\mathcal{K}_{\text{req}} = \{k \mid u_{k,t}=1\}$ exceeds $C$, a subset of $|\mathcal{K}_{\text{req}}| - C$ packets is dropped due to congestion. Transmitted packets are subject to a block error rate $P_{\text{bler}}$.

The receiver reconstructs the haptic signal $\hat{\mathbf{y}}_{k,t}$ based on the result of the transmission:
\begin{equation}
    \hat{\mathbf{y}}_{k,t} = 
    \begin{cases} 
        \mathbf{y}_{k,t}, & \text{if Success}, \\
        \hat{\mathbf{y}}_{k,t-1}, & \text{if } f_{\text{JND}}=0 \text{ Replay}, \\
        \boldsymbol{\mu}_{k,t}, & \text{if } f_{\text{AI}}=0 \text{ or Error},
    \end{cases}
\end{equation}
where $\boldsymbol{\mu}_{k,t}$ is the VarxHAP point prediction used as a fallback recovery mechanism.
\begin{figure}[ht]
    \centering
    \includegraphics[width=0.8\columnwidth]{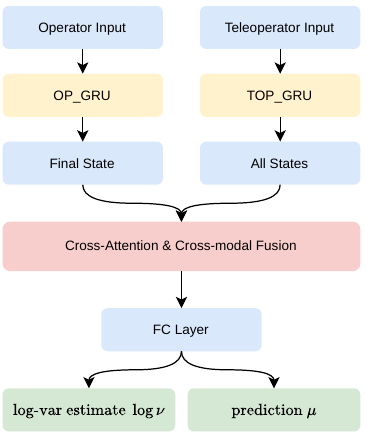}
    \caption{VarxHAP architecture diagram.}
    \label{fig:VarxHAP} 
\end{figure}

\section{VarxHAP}
\label{sec:varxhap}
The VarxHAP framework attempts to generate the true conditional distribution $P(\mathbf{y}_{k,t+m} | \mathbf{H}_{k,t}, \mathbf{O}_{k,t+m})$ to predict the future haptic state $\mathbf{y}_{k,t+m}$ (i.e., 3D force) at $m$ steps ahead. This prediction is conditioned on two physical inputs: $\mathbf{H}_{k,t}$, the recent history of acknowledged 3D force feedback, and $\mathbf{O}_{k,t+m}$, the operator's current 3D position and velocity commands.

\begin{table}[ht]
  \caption{Simulation, Controller, and Model Configuration}
  \label{tab:sim_params}
  \centering
  \renewcommand{\arraystretch}{1.15}
  \resizebox{\columnwidth}{!}{%
  \begin{tabular}{ll|ll}
    \hline
    \multicolumn{2}{c|}{\textbf{Simulation Setup}} & \multicolumn{2}{c}{\textbf{Controller Parameters}} \\
    \textbf{Parameter} & \textbf{Value} & \textbf{Parameter} & \textbf{Value} \\
    \hline
    $K$ (Users) & 60 & $R^*$ (Target Reliability) & $99.999\%$ \\
    $P_{per}$ (PER) & $10^{-3}$ & $K_p$ (Prop. Gain) & $1000.0$ \\
    $\delta$ (JND Threshold) & $10^{-1}$ & $K_i$ (Integral Gain) & $0.01$ \\
    $N_{\text{RB}}$ (Resource Blocks) & 20 & $\eta$ (Base Weight) & $5000.0$ \\
    $T_{sim}$ (Total Steps) & $500,000$ & $\lambda_{min}, \lambda_{max}$ & $[0.1, 2.0]$ \\
    $T_{warmup}$ (Warmup) & $100,000$ & Update Interval & 2000 steps \\
    \hline
    \multicolumn{4}{c}{\textbf{VarxHap Model Hyperparameters}} \\
    \hline
    Operator Input Dim & 6 (Pos+Vel) & Teleop. Input Dim & 3 (Force) \\
    GRU Hidden Size & 128 & Cross-Attn Heads & 4 \\
    GRU Layers & 1 & Attn Key Dimension & 32 \\
    Embedding Dim ($\mathbf{E}_m$) & 32 & FC Layer Size & 64 \\
    Loss Weight $\omega_{nll}$ & 0.1 & Loss Weight $\omega_{cal}$ & 0.05 \\
    Learning Rate & $10^{-3}$ & Weight Decay & $10^{-4}$ \\
    Batch Size & 128 & Dropout Rate & 0.2 \\
    \hline
  \end{tabular}%
  }
\end{table}

\subsection{Architecture and Step-Aware Embedding}
As shown in Figure~\ref{fig:VarxHAP}, the model uses a dual-branch Gated Recurrent Unit (GRU) with cross-attention to fuse historical teleoperator force feedback with up-to-date operator command signals of position and velocity~\cite{kokkinisxHAP}. The model hyperparameters are detailed in Table~\ref{tab:sim_params}.

Standard RNNs are not able to track rollout depths in high-frequency haptics due to negligible inter-sample variance \cite{Tancik2020}. To provide an explicit sense of the prediction horizon, we introduce Step-Aware Embedding $\mathbf{E}_m$, which maps each estimation step $m$ to a high-dimensional space using fixed, multi-frequency sinusoidal functions \cite{Vaswani2017}.

\subsection{Composite Uncertainty Loss}
To ensure perceptual relevance, we mask all losses using an indicator $M_t$. Let $\mathbf{r}_t = \mathbf{y}_t - \boldsymbol{\mu}_t$ denote the prediction residual vector. The masking indicator is defined as:
\begin{equation}
    M_t = \begin{cases} 
        1, & \text{if } \|\mathbf{y}_t\|_2 > \delta_f, \\ 
        0, & \text{otherwise}, 
    \end{cases}
\end{equation}
where $\delta_f$ is a noise floor threshold. With this mask, we only train on data containing significant force activity, which improves inference stability.

The loss term $\mathcal{L}_{\text{pred}}$ for point prediction error minimization is defined as $\mathcal{L}_{\text{pred}} =  \|\mathbf{r}_t\|_2^2 \cdot M_t.$
Although the Mean Squared Error (MSE) term optimizes point prediction accuracy, it does not ensure uncertainty convergence. To capture this, we employ a Gaussian Negative Log-Likelihood (NLL) term $\mathcal{L}_{\text{nll}}$ that penalizes the model for having low uncertainty during high-error events. This is calculated over all dimensions:
\begin{equation}
        \mathcal{L}_{\text{nll}} = \frac{1}{2} \sum_{d=1}^3 \left( \frac{r_{t,d}^2}{v_{t,d}} + \log v_{t,d} \right) \cdot M_t.
\end{equation}
    
NLL maximization alone does not guarantee that the magnitude of uncertainty matches the reconstruction error, which is critical for threshold-based gating. To address this, we introduce a calibration term $\mathcal{L}_{\text{cal}}$ that explicitly regresses the predicted uncertainty towards the observed squared residual, forcing the variance $\boldsymbol{v}_t$ to serve as a direct proxy for error magnitude:
\begin{equation}
        \mathcal{L}_{\text{cal}} = \sum_{d=1}^3 \left( r_{t,d}^2 - v_{t,d} \right)^2 \cdot M_t.
\end{equation}

The total loss function of the VarxHAP model is defined as:
\begin{equation}
    \mathcal{L} = \mathcal{L}_{\text{pred}} + \left( \omega_{\text{nll}} \mathcal{L}_{\text{nll}} + \omega_{\text{cal}} \mathcal{L}_{\text{cal}} \right),
\end{equation}
where $\omega_{\text{nll}}$ and $\omega_{\text{cal}}$ are the weighting coefficients.
\begin{figure}[ht]
    \centering
    \includegraphics[width=\columnwidth]{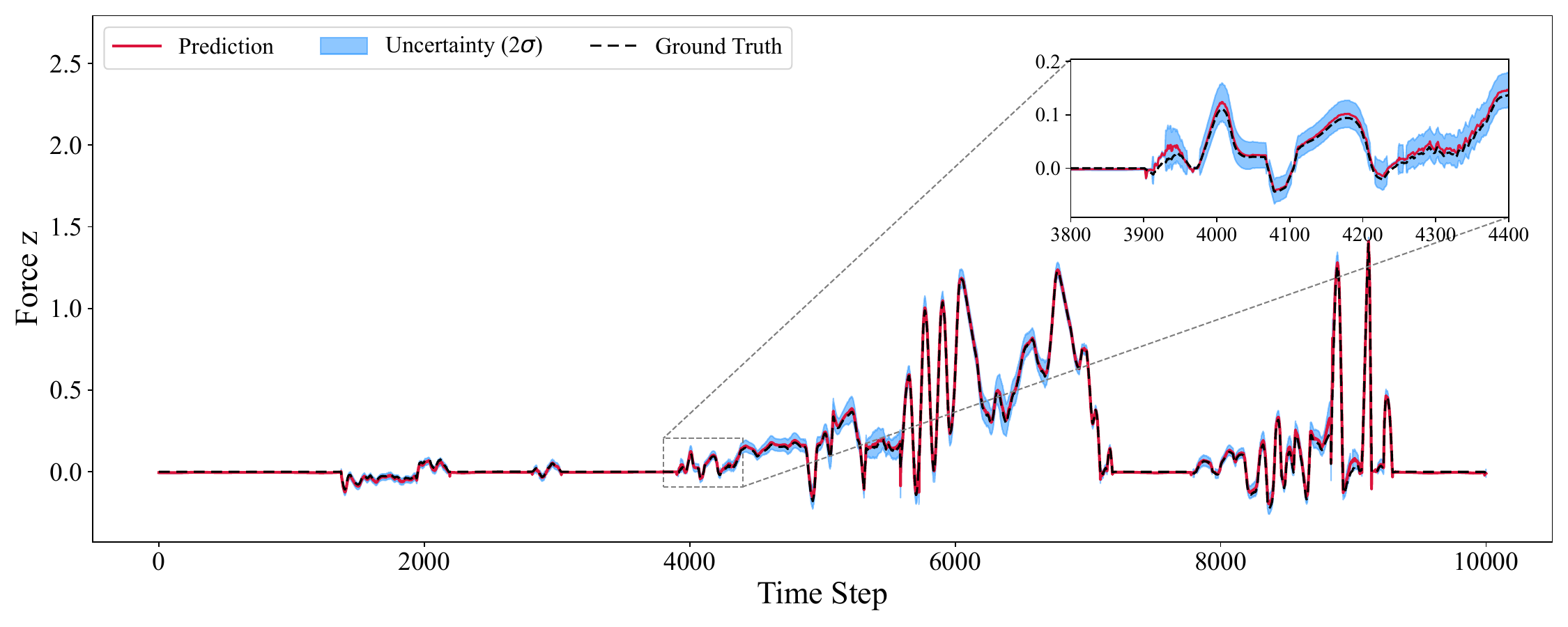}
    \caption{Uncertainty and prediction plot of VarxHAP output over a haptic data trace.}
    \label{fig:var_out} 
\end{figure}

\begin{figure}[ht]
    \centering
    \includegraphics[width=\columnwidth]{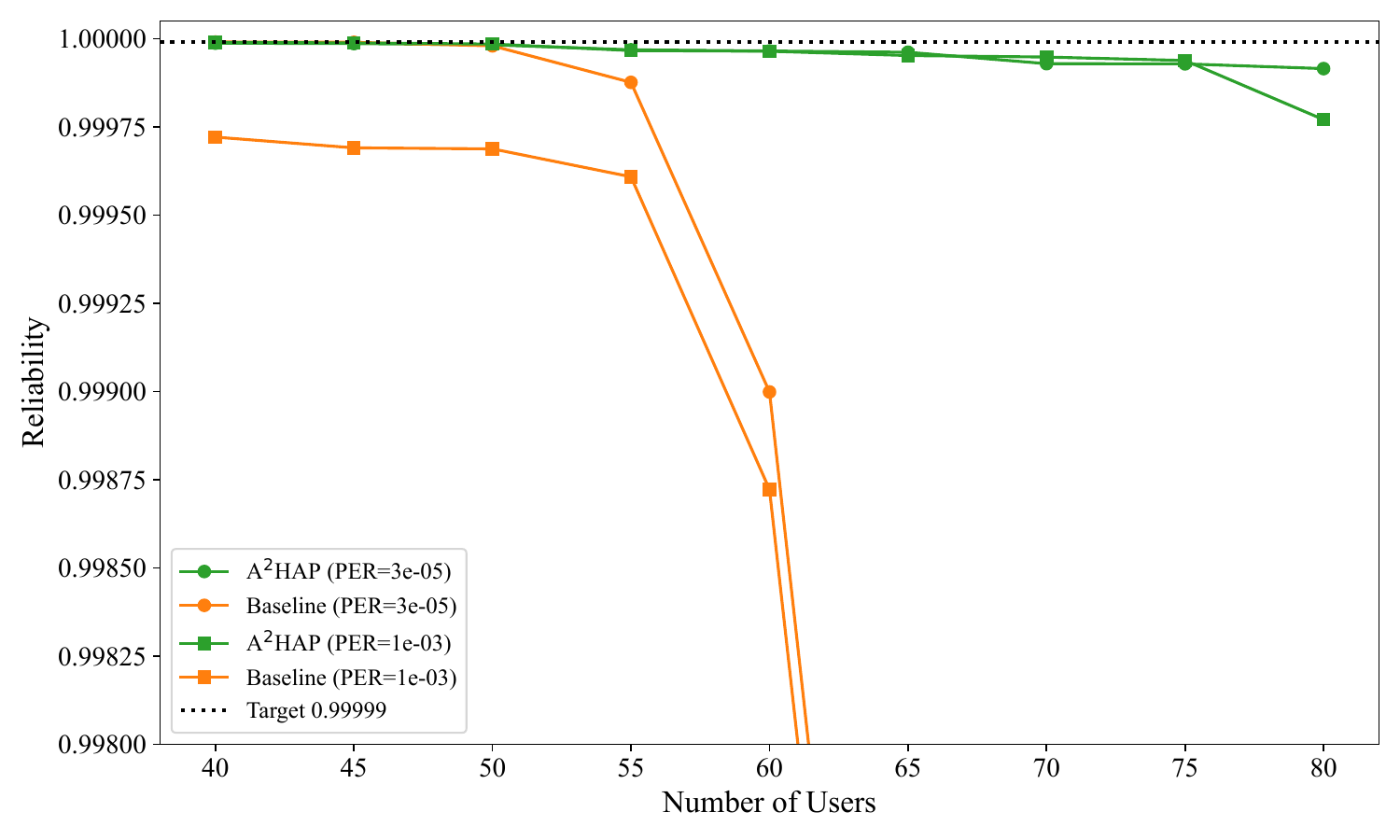}
    \caption{Reliability comparison between baselines and the proposed framework for two different PER values over an increasing number of users.}
    \label{fig:per_comp} 
\end{figure}

\section{Simulation Results}
\label{sec:results}

We evaluate the model in a simulated environment with variable users and round-robin Resource Block (RB) allocation. Transmissions include a statistical Packet Error Rate (PER). Due to the small size of haptic packets, resources are measured in RBs rather than bandwidth. Configuration details are provided in Table~\ref{tab:sim_params}.

\subsection{Dataset and Environment Setup}
While end-to-end teleoperation spans multiple systems, our simulation isolates the Radio Access Network (RAN) edge, as this volatile wireless link is the critical bottleneck for evaluating uncertainty-aware packet suppression. Although static environments typically experience lower error rates, we include a PER of $10^{-3}$ to stress-test the error-resilient controller under adverse conditions, such as deep channel fading or edge-of-cell interference. Furthermore, as demonstrated in Figure~\ref{fig:per_comp}, the framework is also evaluated at a more conservative PER of $10^{-5}$. Experiments utilized real-world 1~kHz haptic traces (100~s per task, with independent train/test sets) recorded via a Geomagic Touch device during distinct teleoperation activities: dynamic object pushing, rigid body interaction, press/hold, and tapping~\cite{Daniel_traces}. Simulations were executed in Python using an Intel 14th Gen CPU and an NVIDIA RTX 4090 GPU.
\begin{figure}[ht]
    \centering
    \includegraphics[width=\columnwidth]{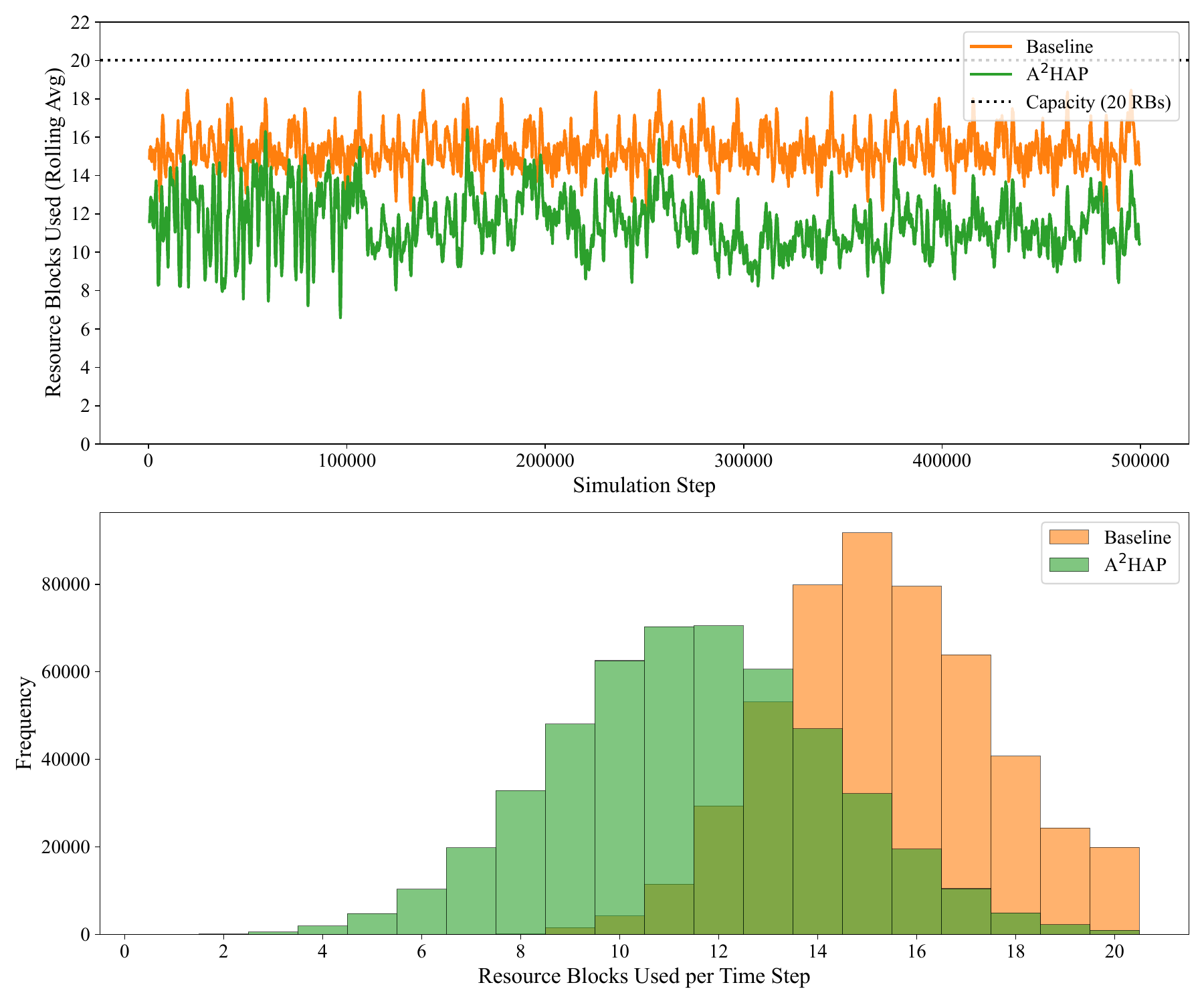}
    \caption{Timeseries and histogram of RB utilization for 60 users.}
    \label{fig:rb_comparison}
\end{figure}

\subsection{Variance Performance Evaluation}
We investigated two training strategies: a sequential two-stage approach (freezing the mean encoder to train the variance head) and a simultaneous end-to-end regime. The two-stage method proved unstable, resulting in gradient explosion during the second phase. Consequently, we adopted simultaneous training for all components.

Figure~\ref{fig:var_out} benchmarks VarxHAP's z-axis force estimation against ground truth, demonstrating high accuracy while assigning high uncertainty to volatile regions. Zoomed sections reveal uncertainty spikes at sharp edges, capturing the unpredictability of rapid force changes. This uncertainty dictates model confidence: low uncertainty enables force packet suppression. During packet loss, however, predictions are used regardless of confidence. Thus, estimation serves a dual purpose: reducing transmission rates and enabling packet recovery.

\subsection{Reliability Increase with A$^2$HAP}
For a reliable estimation to occur in our framework, we define a strict success condition where the estimated force value must not deviate more than $0.1$~N from the true value. We evaluate the performance of A$^2$HAP against a JND-only baseline in Figure~\ref{fig:per_comp} for two different PER values. The proposed framework maintains higher reliability at conservative PER levels, whereas the baseline requires a stringent PER of approximately $3\times10^{-5}$, which is often unattainable in volatile wireless environments. Even when the baseline meets reliability targets under low PER, it consumes significantly more resources. 

In Figure~\ref{fig:rb_comparison}, we demonstrate the RB usage during the simulation for 60 UEs. We observe that A$^2$HAP is more efficient, using 11.5 RBs per Transmission Time Interval (TTI) on average, whereas the baseline requires 15 RBs. This indicates a 25\% reduction in total RB consumption through intelligent packet suppression. In the case of extreme network congestion where all users are highly active, A$^2$HAP utilizes the full resource capacity, but the controller relaxes the confidence threshold to increase the number of estimations, effectively mitigating congestion errors.

\begin{figure}[ht]
    \centering
    \includegraphics[width=\columnwidth]{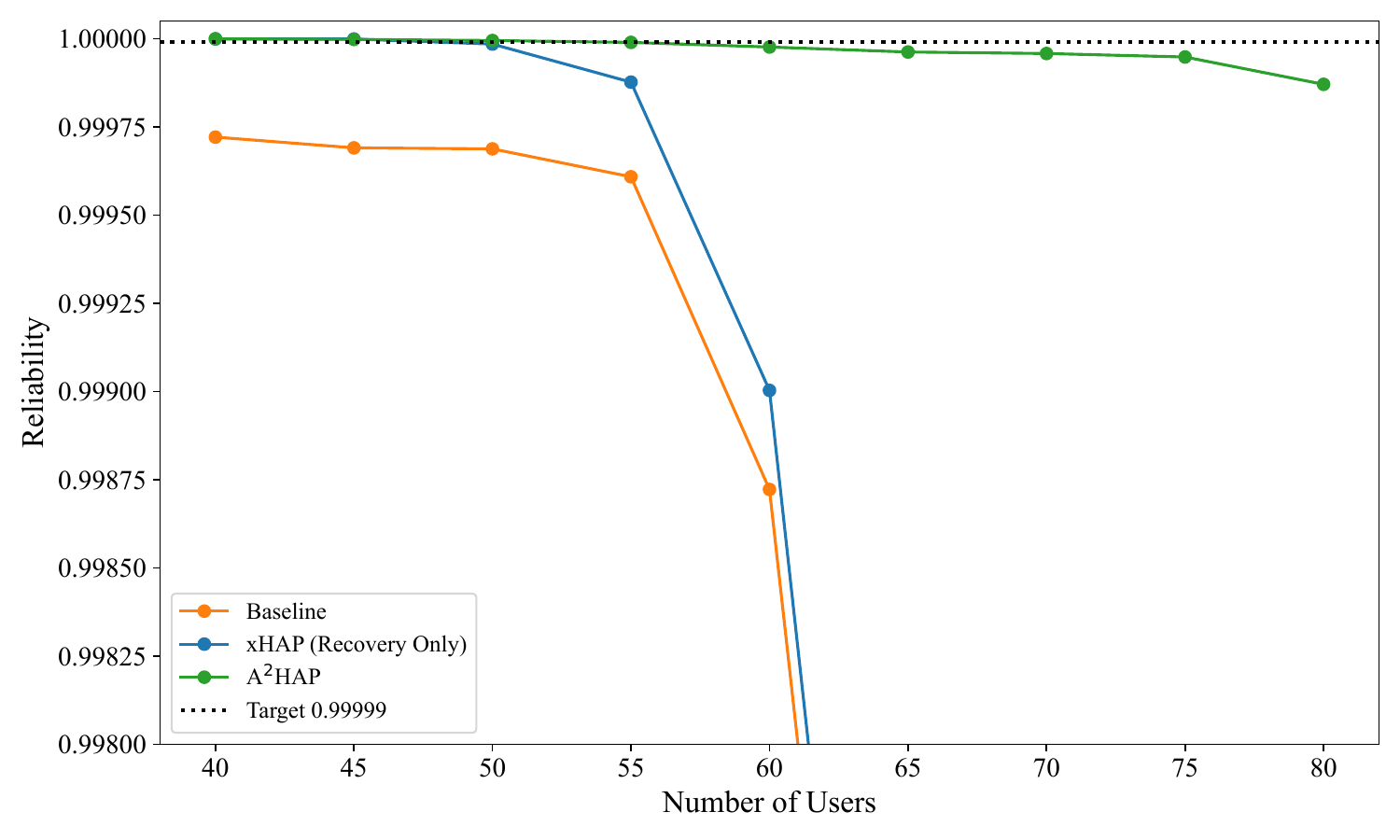}
    \caption{Reliability comparison of VarxHAP vs. xHAP and JND baselines under congestion.}
    \label{fig:all_3_models}
\end{figure}

\begin{figure}[ht]
    \centering
    \includegraphics[width=\columnwidth]{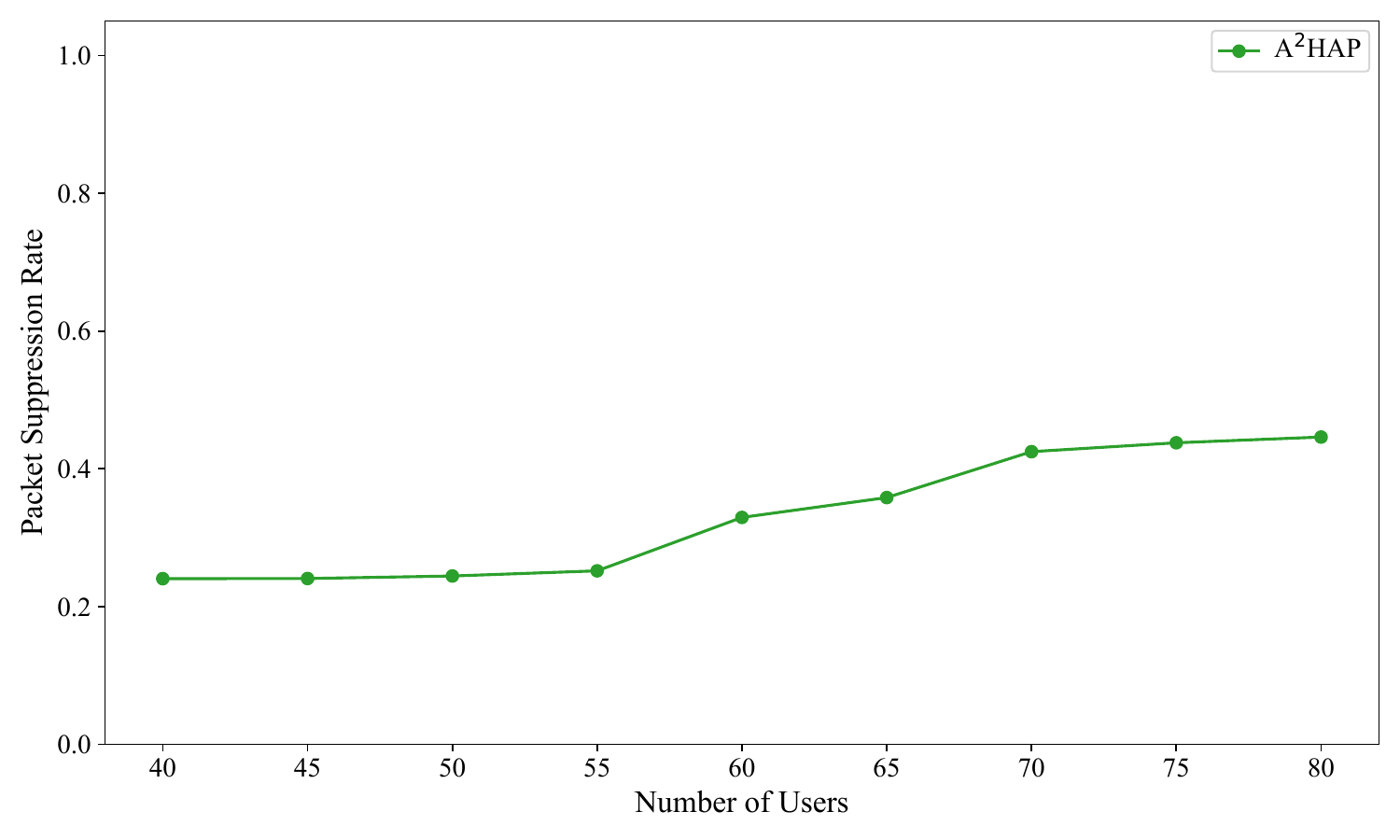}
    \caption{Packet suppression rate of the system as the number of users increases.}
    \label{fig:all_3_suppression}
\end{figure}

Figure~\ref{fig:all_3_models} compares the reliability of the proposed framework against the xHAP recovery method from our previous work and the JND-only baseline. A$^2$HAP accommodates up to 60 users while strictly maintaining the target reliability $R^*$. This represents a 20\% increase in capacity compared to the xHAP model, which is limited to 50 reliable users. The results clearly show that existing methods suffer from rapid performance degradation when congestion occurs. This stability is achieved by the error-resilient controller, which adjusts the confidence level to balance congestion-induced errors against estimation errors. 

As the number of users increases, the controller relaxes the confidence threshold to increase estimation and reduce transmissions, as presented in Figure~\ref{fig:all_3_suppression}. As shown, with VarxHAP estimations, A$^2$HAP can reduce the transmission rate by up to 45\% in high-traffic situations as a trade-off for a minimal drop in reliability. Moreover, the proposed framework yields a smoother, more graceful degradation in reliability than competing methods. This stability stems from the estimator's ability to maintain high-fidelity estimations while significantly reducing the packet rate when network resources are constrained. Essentially, the A$^2$HAP framework achieves a dual advantage: it enhances resource efficiency during low traffic and stabilizes reliability during high traffic. This combined gain is critical for maintaining stable haptic communication in the Tactile Internet.
\section{Conclusion}
\label{sec:conclusion}

This paper introduced A$^2$HAP, a framework that transforms haptic communication from a reactive paradigm to a predictive, uncertainty-aware architecture. This probabilistic approach provides a calibrated confidence metric that serves as a robust control variable for network resource management, bridging the gap between AI-native estimation and the stringent requirements of URLLC. Our experimental results demonstrate that A$^2$HAP accommodates 20\% more users compared to competing methods, while maintaining a $99.999\%$ reliability target. VarxHAP utilizes its internal uncertainty to suppress an additional 25\%-45\% of transmissions. An error-resilient PI controller safeguards this suppression by dynamically adjusting uncertainty thresholds based on network congestion. Future work will extend this probabilistic approach to remote-side joint estimation of position and velocity. Ultimately, integrating uncertainty into the transmission loop is a fundamental enabler for ultra-reliable Tactile Internet communication.
\section*{Acknowledgment}

This research was supported by the TOAST project, funded by the European Union’s Horizon Europe research and innovation program under the Marie Skłodowska-Curie Actions Doctoral Network (Grant Agreement No. 101073465), the Danish Council for Independent Research project eTouch (Grant No. 1127- 00339B) and NordForsk Nordic University Cooperation on Edge Intelligence (Grant No. 168043).

\renewcommand{\baselinestretch}{0.95} 

\bibliographystyle{IEEEtran}
\bibliography{references}

\end{document}